\documentclass[10pt,a4paper,twocolumn,english,pra,aps,showpacs,floatfix,groupedaddress,superscriptaddress]{revtex4-1}

\usepackage{graphicx}
\usepackage{epsfig}
\usepackage[english]{babel}
\usepackage{amsmath}
\usepackage{amssymb}
\usepackage{amsfonts}
\usepackage{longtable}
\usepackage{dcolumn}
\usepackage{bm}
\usepackage{bbm}
\usepackage{color,array}
\usepackage{colortbl}

\usepackage[parfill]{parskip} 

\usepackage{dsfont} 
\usepackage{ctable}
\usepackage{titlesec}

\def\tcr#1{\textcolor{black}{#1}}
\def\tcb#1{\textcolor{black}{#1}}

\begin{document}
\title{{Low temperature thermodynamics of multi-flavored hardcore bosons by the Br\"uckner approach}}

\author{Benedikt Fauseweh}
\email{benedikt.fauseweh@tu-dortmund.de}
\affiliation{Lehrstuhl f\"{u}r Theoretische Physik I, Technische Universit\"at Dortmund, Otto-Hahn Stra\ss{}e 4, 44221 Dortmund, Germany}

\author{G\"otz S.\ Uhrig}
\email{goetz.uhrig@tu-dortmund.de}
\affiliation{Lehrstuhl f\"{u}r Theoretische Physik I, Technische Universit\"at Dortmund, Otto-Hahn Stra\ss{}e 4, 44221 Dortmund, Germany}

\date{\rm\today}

\begin{abstract}
Dynamic correlations of quantum magnets provide valuable information. But especially at finite temperature, many observations, e.g., by inelastic neutron scattering experiments, are not yet quantitatively understood. Generically, the elementary excitations of gapped quantum magnets are hardcore bosons because no two of them can occupy the same site.
The previously introduced diagrammatic Br\"uckner approach dealt with 
hardcore bosons of a single flavor at finite temperature. 
Here, this approach is extended to hardcore bosons of several kinds (flavors). The approach based on ladder diagrams is gauged with so far
unknown rigorous results for the thermal occupation function of multi-flavored hardcore bosons in one dimension with nearest-neighbor hopping.
For low temperatures, it works very well. Furthermore, we study
to which extent the approach is a conserving approximation.
Empirical evidence shows that this is true only in the single-flavor
case.
\end{abstract}


\maketitle

\section{Introduction}

\subsection{General Context}

Understanding the finite temperature effects of correlated quantum systems is a complicated problem. Even at zero temperature the complex interplay of different interactions can lead to a wide range of interesting phenomena, such as bound states,  fractionalization, quantum phase transitions (QPTs) or Bose-Einstein-Condensation \cite{giama08}.
Even more, at finite temperature, interesting features emerge from the interplay of thermal and quantum fluctuations. Especially dynamic correlations are directly affected by microscopic interactions as well as by thermal effects.

From a theoretical point of view, it is often advantageous to describe the 
low-energy physics of quantum systems by an effective model which includes the motion and interactions of quasi-particles of which
the number is conserved or almost conserved. There are numerous ways
to derive effective models, an incomplete list comprises renormalization
approaches \cite{solyo79,metzn12}, continuous unitary transformations (CUTs)
\cite{wegne94,glaze93,glaze94,knett00a,knett03a,kehre06,fisch10a}, and variational
approaches, see for instance Refs.\ \cite{shast81a,uhrig99} for variational
approaches based essentially on intuition and Refs.\ 
\cite{haege13a,keim15a,vande15} for variational
approaches based on matrix product states (MPS).

Variational approaches and CUTs generically provide effective models
at zero temperature. Thus they do not address the thermal effects directly
which renders quantitative comparisons with experiments difficult. 
But if the effective model is formulated in second quantization
it can be evaluated also at finite temperatures by additional
theoretical tools. Such tools are for instance exact diagonalization,
thermodynamic Bethe ansatz, or diagrammatic approaches.

Hardcore bosons are one generic type of elementary excitations in quantum 
lattice systems \cite{knett03a}. 
They behave like normal bosons on different lattice sites, 
but at one site only a single excitation is allowed at maximum. In contrast to free bosons or free fermions, the thermal properties of hardcore bosons are unknown and not trivial even if they are not subject to further interactions. 

The motivation to study hardcore bosonic excitations originates from quantum magnets \cite{sachd90,matsu02}. A variety of elementary hardcore bosonic excitations exist in quantum magnets, e.g., spin-flip excitations in polarized systems, domain walls in
one-dimensional systems with degenerate ground states, or triplon excitations in strongly dimerized systems. Experimentally, inelastic neutron scattering (INS) provides a convenient tool do study the effects of finite temperatures 
on dynamic correlations \cite{loves87}, namely by probing the spin dynamic structure factor (DSF)
\begin{align} 
\label{eq.DSF_definition}
S^{\alpha}(\omega, Q) &= \frac{1}{L} \int_{-\infty}^{\infty} \frac{\mathrm{d} t}{2 \pi} \sum\limits_{l, l'} e^{i \omega t} e^{-i Q(l-l')} \langle S_l^\alpha(t) S_{l'}^\alpha \rangle 
\end{align}
where $L$ is the number of sites.

In this article we  focus on gapped hardcore boson systems. Quantum 
antiferromagnets built from dimers of $S=1/2$ spins on the lattice are of special interest. Examples include spin chains \cite{knett00a,cavad00,tenna12,halg15}, spin ladders \cite{sachd90,schmi05b,exius10,norma11}, coupled ladders 
\cite{uhrig98c,uhrig04a,nafra11,fisch11a} and dimers coupled in several spatial directions \cite{matsu02,ruegg05,quint12,jense14,canev15}.
This underlines that the topic of the present article is of current 
interest and relevant to experiments although it focuses on conceptual issues and benchmarks of the approach.

For gapped quantum magnets the generic DSF spectrum consists of a dominant single quasi-particle peak and weaker multi-particle continua. At finite temperature, the single-particle peak is broadened due to thermal scattering. In recent years, the energy resolution of INS experiments improved significantly. This makes it possible to investigate the position and the shape of the response as function of temperature.

Various theoretical approaches have been applied to describe the effects of finite temperature on the dynamics in quantum magnets. One of them is complete 
diagonalization to compute the Lehmann representation of the DSF 
\cite{fabri97a,mikes06}. For one-dimensional systems, Essler and co-workers 
developed a variety of approaches based on integrability, Bethe-Ansatz equations and fermionization in Refs.\ \cite{essle08,james08,essle09,goetz10}. Certain
thermal effects such as the  narrowing of the dispersion and shifts
of spectral weight can already be captured by mean-field approaches \cite{exius10}. 
The combination of integrability and density-matrix renormalization
is a very useful strategy where applicable \cite{lake13}. An
advanced numerical approach is based on density-matrix renormalization
formulated in matrix product states in the frequency space \cite{tiege14}. 

The above listed methods rely essentially on the one-dimensionality of the system under study. In principle, the complete diagonalization can be applied to any dimension. But the tractable system size measured in the linear extension becomes very small for dimensions higher than one.
Thus, recent finite temperature experiments on 3D materials \cite{quint12} can not be described by these approaches. A $1/z$ expansion with $z$ being the coordination number by Jensen \cite{jense14} agrees remarkably with the experimental data. The underlying idea relies  on a description of the scattering  of the elementary excitations with an effective medium  \cite{jense84,jense94,jense11}.
The caveat of the approach is that it is unclear how it can be systematically extended beyond the linear order in $1/z$. Already in this linear order, non-physical double poles have to be eliminated by a plausible assumption in order to preserve causality.

\subsection{Diagrammatic Approach}

In 2014, we introduced a  diagrammatic perturbative approach based on a low-temperature expansion which does not require a particular dimensionality \cite{fause14}. The systematic control parameter of the approach is the low density of thermally excited hardcore bosons
which is proportional to $\exp(-\beta \Delta)$ in a gapped system where $\beta$ is inverse temperature and $\Delta$  the energy gap.

The method works conceptually in any dimension; it directly addresses
 frequency and momentum space in the thermodynamic
limit. Therefore finite size effects and the ill-defined continuation 
from imaginary Matsubara frequencies to real frequencies are avoided. 
Furthermore the approach lends itself for effective models at $T=0$
derived by other methods.

A key ingredient for the applicability of the advocated approach is 
that the system conserves the number of quasi-particles (elementary
excitations), i.e., there are no finite anomalous Green functions and the vacuum is the ground state of the system. 

\tcb{One may think that this property contradicts the existence of quantum fluctuations in the original model. But this is \emph{not} the case. The idea is to treat the quantum fluctuations beforehand by another approach,
based on variation or on unitary transformations. The effective model
obtained should  conserve the number of quasi-particles with the
ground state being their vacuum. \emph{On top} of this effective Hamiltonian, we apply the Br\"uckner approach to treat the thermal effects in the spectrum.}

The method is motivated by the since long known Br\"uckner theory in nuclear matter \cite{fette71} \tcr{and is discussed as the T-matrix approximation in Ref.\ \onlinecite{kadanoffbaym62}}.
But it has also been employed for spin systems at zero temperature 
by Sushkov and co-workers \cite{kotov98,sushk98}. In these
applications, however, the quantum fluctuations were
treated diagrammatically  since the Br\"uckner idea
was not applied to a particle-conserving model \cite{kotov99a}
which makes its application non trivial even at zero temperature.

The  Br\"uckner approach takes into account that the elementary excitations in quantum paramagnets are hardcore bosons, i.e., a site can be occupied at maximum by a single excitation. This is enforced
by an infinite on-site repulsion.
Note that at zero temperature the hardcore constraint is irrelevant for the single-particle dynamics if the Hamiltonian is particle conserving.
At finite temperature, however, the hardcore scattering provides the primary source for the loss of coherence of the elementary excitations. Then, the single particle is scattered from the thermally excited background of particles. 

In a first article \cite{fause14} we introduced the 
Br\"uckner approach at finite temperatures and tested it for hardcore
bosons with a single flavor in one dimension. The advantage of
this model was that it can be solved using the Jordan-Wigner transformation \cite{jorda28}. It turned out that the Br\"uckner approach based on ladder diagrams
 works very well. In particular, we empirically found that the intricate sum rules of hardcore bosons were conserved even at high temperatures at and 
beyond the energy gap  $\Delta$ within numerical accuracy. Furthermore, the approach yielded
a very good approximation for the exact thermal occupation for all temperatures.
This has been recently confirmed and quantified even down to vanishing gaps 
\cite{strei15}. The asymmetric line-shape broadening and 
bandwidth narrowing found in other theoretical studies 
\cite{mikes06,essle08,james08,essle09,goetz10} and in
experimental studies, e.g., in Ref.\ \cite{tenna12}, are retrieved.

In the present paper, we extend the finite-temperature Br\"uckner approach to systems with multiple flavors, i.e.,  there are hardcore bosons of different kinds.  But each single site can still only be occupied by a single boson. This situation is relevant wherever
the local site is represented by a more complex
quantum system with a Hilbert space dimension larger than two.
A generic example are  systems built from coupled dimers,
each consisting of two spins $S=1/2$. In these systems the elementary excitations are mobile $S=1$ triplets, which we call triplons 
to distinguish them from magnons in ordered magnets
\cite{schmi03c}.  For full SU(2) symmetry the triplons are threefold degenerate, i.e., they are hardcore bosons with three flavors.  

The goal of our study is to investigate the multi-flavor extension of the 
Br\"uckner approach and to gauge the method against exact results. Note
that a benchmarked method to calculate finite temperature correlations applicable in any dimension is essential to describe and to explain experimental observations in quantum magnets. We study the performance of the method with a special focus on sum rules and discuss to which extent our approach
represents a conserving approximation for hardcore particles. 

We stay here with the one-dimensional case
because it allows us to compare the diagrammatic approach to rigorous
results. But it must be emphasized that there is no conceptual obstacle
for the application of the diagrammatic approach in higher dimensions
as well.

In order to have rigorous results at our disposal we derive the thermal occupation function for hardcore bosons in one dimension exactly for a nearest neighbor hopping.

The paper is organized as follows: In Section \ref{sec.model} we 
introduce the multi-flavor hardcore boson model and discuss some of its 
properties. In the following Section \ref{sec.thermal_occupation} we 
derive the exact thermal occupation for a special case of 
hardcore bosons in one dimension with nearest-neighbor
hopping. In Section \ref{sec.diagramm} we extend
 the Br\"uckner approach on the basis of ladder diagrams 
to the multi-flavor case.  Next, in Section 
\ref{sec.results} we evaluate the results of the diagrammatic Br\"uckner approach on the basis of the findings of the preceding sections. In
 Section \ref{sec.finite_U} we perform various calculations at finite repulsion,
 i.e., the bosons repel each other only with a finite energy cost, to clarify to which extent conserving approximations
exist for hardcore bosons. Finally we conclude in Section 
\ref{sec.conclusion}.

\section{Model and Exact Results}

\subsection{General Properties}
\label{sec.model}

In this section, we introduce the multi-flavor hardcore boson model and discuss some of its properties such as commutators, spectral functions, and sum rules.

The general Hamiltonian of conserved hardcore bosons reads
\begin{align} 
\label{eq.model_ham}
H_0 &= E_0 + \sum\limits_{i,d,\alpha} \left( w_d^{\alpha} b_{i,\alpha}^\dagger b_{i+d,\alpha}^{\phantom\dagger} + \mathrm{h.c.} \right) 
\nonumber  \\ &+
\hspace{-0.2cm}
\sum\limits_{\tiny 
\begin{matrix} i,  d_1,  d_2,  d_3  \\
\alpha,  \phi,  \gamma,  \xi \end{matrix}} V_{d_1,d_2,d_3}^{\alpha,\phi,\gamma,\xi} b_{i,\alpha}^\dagger b_{i+d_1,\phi}^\dagger  b_{i+d_2,\gamma}^{\phantom\dagger} b_{i+d_3,\xi}^{\phantom\dagger} + \dots,
\end{align}
where the dots stand for possible higher particle-number interactions. 
Here $E_0$ denotes the ground state energy, $\lbrace i,d,d_1,d_2,d_3 \rbrace$ denote site indices which need not be restricted to one dimension, $\lbrace \alpha, \phi, \gamma, \xi \rbrace$ denote flavor 
indices, $w_d^{\alpha,\phi}$ are hopping matrix elements, $V_{d_1,d_2,d_3}^{\alpha,\phi,\gamma,\xi}$ are interaction vertices and $b_{i,\alpha}^\dagger, b_{i,\alpha}^{\phantom\dagger}$ denote hardcore bosonic creation and annihilation operators. The hopping matrix elements must fulfill the hardcore constraint, i.e., $d \neq 0, d_1 \neq 0, d_2 \neq d_3$.

The model \eqref{eq.model_ham} does not include terms that change the number of particles. These terms would induce additional quantum fluctuations and the vacuum would not be the ground state of the system. It is understood that particle number violating terms have already been accounted for by other methods, e.g., based on 
unitary transformations or on variational principles.
\tcb{Hence we \emph{do not neglect} quantum fluctuations, but the Hamiltonian 
\eqref{eq.model_ham} describes the hopping and the interactions of conserved effective particles. We aim at the systematic description of thermal effects in spectral properties for the above effective Hamiltonian.}

In Ref.\  \cite{fause14} we considered the single-flavor case of 
a chain with nearest-neighbor hopping without any additional interactions. Here, we address the consequences of introducing several flavors.

The hardcore operators in second quantization fulfill the algebra,
\begin{align}
\label{eq.triplon_realspace}
\left[ b_{i,\alpha}^{\phantom\dagger}, b_{j,\phi}^\dagger \right] = \delta_{ij} \left( \delta_{\alpha \phi} \left( 1 - \sum\limits_\gamma b_{i,\gamma}^\dagger b_{i,\gamma}^{\phantom\dagger} \right) - b_{i,\phi}^\dagger b_{i,\alpha}^{\phantom\dagger} \right) .
\end{align}
Transforming the hardcore commutator into Fourier space yields
\begin{align}
\label{eq.triplon_fourier}
\left[ b_{k,\alpha}, b^\dagger_{k', \phi} \right] &= \frac{1}{L} \sum\limits_{q}  \left( \delta_{\alpha,\phi} \left( 1- \sum\limits_{\gamma} b^\dagger_{k'+q,\gamma} b_{k+q, \gamma} \right)  \right. \nonumber \\
 &- \left. \phantom{\sum\limits_{q}} \hspace{-0.5cm} b^\dagger_{k'+q,\phi} b_{k+q, \alpha}  \right) .
\end{align}

The quantity of interest is the single-particle spectral function depending
on momentum $p$ and frequency $\omega$
\begin{align}
\nonumber
A^{\alpha}(p, \omega) &= \frac{-1}{\pi} \lim\limits_{i \omega_\nu \rightarrow \omega + i 0^+} 
\\
& \quad \mathrm{Im} \int\limits_0^\beta \mathrm{d}\tau  e^{i \omega_\nu \tau} \frac{1}{\sqrt{L}} \sum\limits_j e^{-i p j} G^{\alpha}(j, \tau) ,
\end{align}
where $G^\alpha(j, \tau)$ is the single-particle temperature Green function
\begin{align}
G^{\alpha}(j, \tau) = - \left\langle T\left\{ b_{j,\alpha}^\dagger(-i\tau) b_{0,\alpha}^{\phantom\dagger}(0) \right\}\right\rangle .
\end{align}
We use the Matsubara frequencies $\omega_\nu = 2\nu \pi / \beta$ and the inverse temperature $\beta = 1/T$. For hardcore bosons the spectral function is a non-trivial quantity because it can not be calculated exactly, even in the absence of any additional interactions beyond the hardcore constraint. The hardcore constraint itself represents
a strong interaction inducing intricate correlations.

The spectral function is related to the dynamic structure factor (DSF) by means of  the fluctuation-dissipation theorem
\begin{align}
\label{eq.fluctuation_dissipation}
S^{\alpha}(p, \omega) = \frac{1}{1-e^{-\beta\omega}} \left[ A^{\alpha}(p, \omega) + A^{\alpha}(p, -\omega) \right].
\end{align}
The DSF is directly accessible by scattering experiments.

Sum rules, i.e., the integrals of the spectral function over frequency, often provide a convenient way to evaluate the accuracy of various approximations. For normal bosons, the integration of the spectral function over frequency yields a temperature independent constant of unity. But for hardcore bosons this is no longer true. The general sum rule states that the equality
\begin{align}
\label{eq.sum_rule_general}
 \int\limits_{-\infty}^\infty A^{\alpha}(p, \omega) \mathrm{d} \omega = \left\langle \left[ b_{p,\alpha}^{\phantom\dagger}, b_{p,\alpha}^\dagger \right] \right\rangle,
\end{align} 
holds. Evaluating Eq.\ \eqref{eq.sum_rule_general} with the commutator in Fourier space \eqref{eq.triplon_fourier} leads to
\begin{align}
\label{eq.multi_flavor_sum_rule}
\int\limits_{-\infty}^\infty A^{\alpha}(p, \omega) \mathrm{d} \omega = 
1 - (1+N_\mathrm{f}) n(T),
\end{align}
where $N_\mathrm{f}$ denotes the number of different flavors, e.g., $N_\mathrm{f} = 3$ in the triplon case.
The thermal occupation $n(T) = \frac{1}{L} \sum_q \langle b_{q,\alpha}^\dagger b_{q,\alpha}^{\phantom\dagger} \rangle$ is a function of temperature and independent of the flavor if the different flavors are degenerate.

At zero temperature, the thermal occupation vanishes and the standard boson sum 
rule is recovered. At infinite temperature, the thermal occupation function yields 
$n(T) = 1/(1+N_\mathrm{f})$, so that the integration of the spectral function over frequency yields zero. 
In between, the thermal occupation  is a non-trivial quantity. 

\subsection{Exact Thermal Occupation in One Dimension}
\label{sec.thermal_occupation}

In this subsection, we calculate the thermal occupation  of 
 multi-flavored hardcore bosons with nearest-neighbor hopping along
a chain. We consider a chain of $L$ sites of hardcore bosons with a nearest neighbor hopping Hamiltonian without any additional interactions
$V$
\begin{align}
\label{eq.hamiltonian}
H &= \sum\limits_{i=1, \alpha}^{i=L} \left(\Delta + \frac{W}{2} \right) 
b_{i,\alpha}^\dagger b_{i,\alpha}^{\phantom\dagger} 
\nonumber
\\
&- \sum\limits_{i=1, \alpha}^{i=L} \frac{W}{4} \left( b_{i,\alpha}^\dagger b_{i+1,\alpha}^{\phantom\dagger} + \mathrm{h.c.} \right).
\end{align}
The energy gap is given by $\Delta>0$ while $W>0$ is the band width of the 
dispersion
\begin{align}
\label{eq.dispersion}
\omega(k) = \Delta + \frac{W}{2} \left[ 1-\cos(k) \right].
\end{align}
Periodic boundary conditions are implicit and 
 $k$ is the momentum of the excitations. The minimum $\Delta$ of the dispersion is found at $k=0$. The ground state of the system is given by the vacuum state. 

Since we only allow nearest neighbor hopping, the bosons cannot skip over each other. This implies that a given sequence of the flavors
along the chain is conserved in the dynamics induced by $H$ in 
\eqref{eq.hamiltonian}. Thus the flavor degree of freedom and the
hopping dynamics can be considered to be completely decoupled
for a fixed number of bosons. The flavor degree adds a certain degree of entropy for given number $N$ of bosons. The hopping of the hardcore bosons acts as if all bosons have the same color.

To calculate the thermodynamics of the flavorless hardcore bosons we apply a Jordan-Wigner transformation
\begin{align}
c_j &= \text{exp}(\pi i \sum_{i<j} b_i^\dagger b_i^{\phantom\dagger}) b_j & c_j^\dagger &= \text{exp}(- \pi i \sum_{i<j} b_i^\dagger b_i^{\phantom\dagger}) b_j^\dagger \nonumber \\
b_j &= \text{exp}(- \pi i \sum_{i<j} c_i^\dagger c_i^{\phantom\dagger}) c_j & b_j^\dagger &= \text{exp}( \pi i \sum_{i<j} c_i^\dagger c_i^{\phantom\dagger}) c_j^\dagger
\end{align}
so that we obtain a fermionic chain model.
Since only nearest neighbor hopping is allowed the Hamiltonian remains bilinear when expressed in fermions.
The grand canonical partition function of these fermions is simple
and given by
\begin{align}
Z_g(\beta, \mu_0) = \mathrm{Tr} \left[ \exp\left(-\beta H_0 + \beta \mu_0 \hat{n}\right) \right]
\end{align}
where $\hat{n} = \sum_j \hat{n}_j$ counts the number of fermions and $\left\langle \hat{n} \right\rangle = N$ is the 
expectation value for the number of thermally excited fermions in the chain. This is the total number of thermally excited bosons
irrespective of their flavor. The Hamiltonian $H_0$ in terms of fermions is given by
\begin{align} 
H_0= \sum\limits_{k} c_{k}^\dagger c_{k}^{\phantom\dagger} 
\omega(k),
\end{align}
where the Fourier transformation to momentum space has already
been carried out to diagonalize the Hamiltonian. 
The grand canonical partition function in the eigen basis of the fermions reads
\begin{align}
Z_g(\beta, \mu_0) &= \prod\limits_{k} \left( 1 + \exp\left(-\beta \left[\omega(k) - \mu_0 \right] \right) \right)
\end{align}
implying the grand canonical potential

\begin{subequations}
\begin{align}
J_0(\beta,\mu_0) &= \frac{-1}{\beta} \ln Z_g(\beta, \mu_0) \\
			 &= \frac{-1}{\beta} \sum\limits_{k} \ln \left[ 1 + \exp\left(-\beta \omega(k) \right) \exp\left(\beta \mu_0 \right) \right].
\end{align}
\end{subequations}

In the grand canonical ensemble, the particle number $N$ is no canonic thermodynamic variable so that we cannot simply add the effect of
different flavors. Hence we perform a Legendre transformation to the 
free energy
\begin{subequations}
\begin{align}
F_0(\beta,N) & = J + \mu_0 N \\
 \quad \mathrm{dF}_0 &= -S \mathrm{d}T + \mu_0 \mathrm{d}N.
\end{align}
which reads
\begin{align}
F_0(\beta, N) &= \frac{-1}{\beta} \sum\limits_{k} \ln \left[ 1 + \exp\left(-\beta \omega(k) \right) \exp\left(\beta \mu_0(N) \right) \right]  \nonumber \\
 & + \mu_0(N) N.
\end{align}
\end{subequations}
At fixed $N$ the effect of the different flavors can
be accounted for. The additional degree of freedom leads to an entropy increase and hence to a decrease in the free energy
\begin{subequations}
\begin{align}
F &= F_0 - T S_\mathrm{flavor} \\
 &=  F_0 - \frac{1}{\beta} \ln \left( N_{\mathrm{f}}^N \right)
\end{align}
\end{subequations}
where the flavor entropy 
$S_\mathrm{flavor}  = \ln \left( N_{\mathrm{f}}^N \right)$ is determined by the number of flavors $N_\mathrm{f}$ and the number of thermal excitations $N$. The free energy $F$ takes the effect of different flavors into account.

In a next step, we calculate the modified chemical potential
\begin{subequations}
\begin{align}
\mu &= \partial_N F(\beta,N) \\
    &= \mu_0 - \underbrace{\frac{1}{\beta} \ln 
		\left( N_{\mathrm{f}}\right)}_{\mu_F(\beta)}.
\end{align}
\end{subequations}
The effect of different flavors is seen in the increase of the effective chemical potential.
A short calculation provides us with the modified grand canonical potential 
$J =  F - \mu N$ from which we finally obtain the thermal occupation
\begin{subequations}
\begin{align}
n_\mathrm{exact}(T) &= \frac{N(T)}{L N_\mathrm{f}}
\\ \label{eq.exact_occupation}
& = \frac{1}{L} \sum\limits_k \left[ \exp\left(\beta \omega(k) \right) +  
N_\mathrm{f} \right]^{-1} .
\end{align}
\end{subequations}
This result shows the correct behaviour for $T=0 \Rightarrow n(T) = 0$ and for $T \rightarrow \infty \Rightarrow n(T) = 1/(1+N_\mathrm{f})$. For $N_\mathrm{f} = 1$ the Fermi-function is recovered as it has to be.

We stress that the mapping to free fermions by the Jordan-Wigner transformation leads to a separation of the flavor and the particle dynamics only for a \emph{fixed} number of particles. If we consider dynamic correlation functions which involve changes of the number of bosons we do not know
of a mapping to a solvable model which makes these correlations tractable.
But further attempts are called for.

\section{Diagrammatic Approach}
\label{sec.diagramm}

In this section, we introduce the diagrammatic approach to treat the 
hardcore repulsion at finite temperatures. The aim is an expansion in the small parameter $\exp(-\beta\Delta)$. 
The approach has been already introduced in \cite{fause14} for 
bosons of a single flavor. Here we focus on the changes necessary to treat multi-flavor hardcore bosons.

There is no diagrammatic perturbation theory for hardcore bosons at finite temperatures. Thus, the key idea is to start from normal bosons and to enforce the hardcore constraint by means of  
an infinite on-site interaction $U\to\infty$.
For nuclear matter and He$^3$ this idea was introduced under the name of Br\"uckner theory \cite{fette71}.
It is important to see the difference of our approach to the one in Refs.\ 
\cite{fette71} and \cite{kotov98,sushk98,kotov99a} where the leading effects of quantum fluctuations are addressed by Br\"uckner theory.  In the present application, the Br\"uckner theory  captures the leading \emph{thermal} 
effects on the spectrum, e.g., 
the thermal broadening of single-particle lines.

We replace the initial Hamiltonian $H_0$ in Eq.\ \eqref{eq.model_ham} by
\begin{subequations}
\begin{align}
H &= H_0 + H_U \\ H_U &= \lim\limits_{U \rightarrow \infty} U 
\sum\limits_{i} \sum\limits_{\alpha, \phi} b_{i,\alpha}^\dagger 
b_{i,\phi}^\dagger b_{i,\phi}^{\phantom\dagger} 
b_{i,\alpha}^{\phantom\dagger} ,
\end{align}
\end{subequations}
where $U$ is the strength of the auxiliary local repulsion.
The main difference to \cite{fause14} is that different flavors
repel each other on the same site as well.

We want to calculate the single particle spectral function 
$A(p,\omega)$. At zero temperature, the spectral function is a 
$\delta$-function located at the energy given by the 
dispersion, i.e., $A(p,\omega) = \delta(\omega- \omega(p))$. At finite temperature the excitations are scattered by the thermally excited background generating a broadening of the spectral function in 
frequency space. 

\begin{figure}[ht]
\centering
\includegraphics[width=0.8\columnwidth]{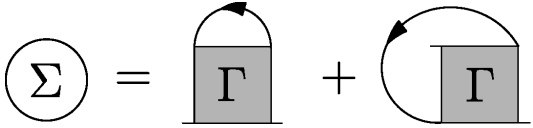}
\caption{Self-energy diagrams in leading order in $\exp(-\beta\Delta)$. The first diagram are Hartree-like diagrams and the second refers to the Fock-like diagrams with a renormalized interaction given by
the scattering amplitude $\Gamma$.}
\label{fig.self_energy}
\end{figure}

\begin{figure}[ht]
\centering
\includegraphics[width=0.99\columnwidth]{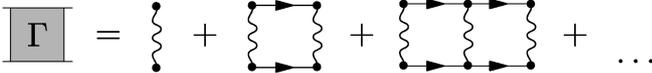}
\caption{Definition of the scattering amplitude $\Gamma$}
\label{fig.ladder2}
\end{figure}

Since we have to take the limit $U \rightarrow \infty$, truncated perturbation theory in the repulsion strength $U$ is not useful. 
Instead, we aim at a low-temperature expansion.
On the diagrammatic level one observes that each propagator 
running backwards in imaginary time implies a factor $\exp(-\beta\Delta)$.
Thus, in first order in $\exp(-\beta\Delta)$ one has to sum all self-energy diagrams with one backward running propagator. 
This amounts up to the summation of the ladder diagrams
as depicted in Figs.\ \ref{fig.self_energy} and \ref{fig.ladder2}.
The gray box indicates the scattering amplitude or 
effective interaction $\Gamma$ given in Fig.\ \ref{fig.ladder2}. 

\begin{figure}[ht]
\centering
\includegraphics[width=0.99\columnwidth]{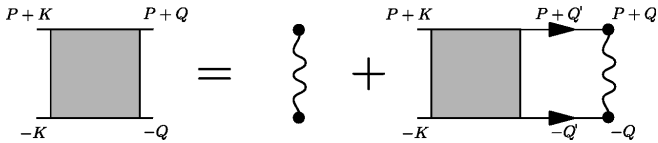}
\caption{Bethe-Salpeter equation for the scattering amplitude}
\label{fig.bethe_salpeter}
\end{figure}

The scattering amplitude fulfills the Bethe-Salpeter equation given in Fig.\ \ref{fig.bethe_salpeter} which is essentially a geometric series.
Due to the simple structure of the local interaction, it can be easily evaluated leading to 
\begin{align}
\label{eq.scattering_amplitude}
\Gamma^{\alpha,\phi}(P) = \frac{\frac{U}{L \beta}}
{1 + \frac{U}{L \beta} \sum\limits_Q G^{\alpha}(P+Q) G^{\phi}(-Q)}.
\end{align}
The capital letters like $P$ and $Q$ denote the 2-moment 
$P = (p, i \omega_p)$ and so on.  In contrast to the situation
in our previous work \cite{fause14} the scattering amplitude depends on two flavor indices $\alpha$ and $\phi$ 
denoting the different flavors of the propagators $G$ in 
Eq.\ \eqref{eq.scattering_amplitude}. The same applies to the spectral functions $\rho_p(x)$ and $\bar{\rho}_p(x)$ introduced in 
Ref.\ \cite{fause14}.

Evaluating the diagrams in Fig.\ \ref{fig.self_energy} yields
the expression
\begin{align}
\label{eq.self_energy1}
\Sigma^\alpha(P) = \frac{1}{L} \sum\limits_\phi \sum\limits_{K} (1+\delta_{\alpha,\phi}) G^\phi(K) \Gamma^{\alpha,\phi}(P+K),
\end{align}
for the self-energy. The term proportional to $1$ stems from the 
Hartree-like diagram whereas the term proportional to $\delta_{\alpha,\phi}$ stems from the Fock-like diagram.
If the flavors are degenerate, i.e., the propagator 
$G^\alpha(P) = G^\phi(P) = G(P)$ does not depend on the index $\alpha$,
one can simply replace the sum over the flavor index $\phi$ in 
\eqref{eq.self_energy1} by a factor
\begin{align}
\label{eq.self_energy2}
\Sigma^\alpha(P) = \frac{1+N_{\mathrm{f}}}{L} \sum\limits_{K} G^\alpha(K) \Gamma^{\alpha,\alpha}(P+K).
\end{align}
Once the self-energy is calculated, one determines the spectral function
\begin{align}
A^\alpha(p, \omega) &= \nonumber\\
& \frac{-1}{\pi} \frac{\mathrm{Im} \Sigma^\alpha(\omega,p)}
{\left(\omega - \omega(p) - \mathrm{Re} \Sigma^\alpha(\omega,p) \right)^2 
+ \left( \mathrm{Im} \Sigma^\alpha(\omega, p) \right)^2} .
\end{align}
We conclude that the main difference to the case
of hardcore bosons of a single flavor is the factor $1+N_\mathrm{f}$ 
in Eq.\ \eqref{eq.self_energy2} for degenerate flavors.

We emphasize that the spectral function is computed self-consistently, i.e., we use the bare propagators 
$G_0^\alpha(P) = 1/(i \omega_p - \omega^\alpha(p))$ only as initial guess
 for the dressed Green function $G^\alpha(P)$ in the diagrams. After the first computation of the self-energy we iterate all following calculations with the full propagator $G^\alpha(P) = 1/(i \omega_p - \omega^\alpha(p) - \Sigma^\alpha(P))$ with the previously obtained 
self-energy until we reach self-consistency within numerical tolerance.

For normal bosons, i.e., without infinite interactions, the
self-consistent approximation based on ladder diagrams is a conserving
approximation as introduced by Kadanoff and Baym \cite{baym61,baym62}. Such approximations are derived from a generating functional $\Phi$ depending
on the dressed propagators. The self-energy diagrams are obtained
from $\Phi$ by functional derivation with respect to the propagators.
This construction ensure that conservation laws and sum rules are
fulfilled. Thus the question arises whether the approach 
is also conserving for hardcore bosons.

\tcr{Kadanoff and Baym showed the applicability of the T-matrix 
approximation to single-flavor bosons with hardcore interactions 
\cite{kadanoffbaym62} in continuum models. 
They also established that the T-matrix
approximation is conserving. To our knowledge, however, they did not
establish generally that the T-matrix approximation is conserving
for infinite interactions.
In Ref.\ \onlinecite{kadanoffbaym62} there is no discussion of the
 difference between first integrating over frequency and then taking the limit $U \rightarrow \infty$ and the opposite sequence.} 

Empirically, we found previously \cite{fause14}
that our approach conserves the sum rules for hardcore bosons of a single flavor
within numerical accuracy. We emphasize that this is a non-trivial statement which 
does not result from continuity in the limit $U\to\infty$.
The integration over all frequencies from $-\infty$ to $\infty$
does not commute with the limit $U\to\infty$ because the latter
shifts weight to infinity. This weight is lost in the integration
explaining why the weight of the spectral functions of hardcore
bosons is reduced as expressed by the right hand side of
Eq.\ \eqref{eq.multi_flavor_sum_rule}.
The other sequence, namely integrating first and then
having $U$ tend to infinity describes normal bosons and
does not alter the sum rule which remains unity.
We will show this explicitly in Section \ref{sec.finite_U}.

\section{Results for Multi-Flavor Bosons}
\label{sec.results}

In this section we discuss the results of the Br\"uckner approach in the multi-flavor case. We focus on the changes upon passing 
from $N_\mathrm{f} = 1$ to $N_\mathrm{f} = 2$ and $N_\mathrm{f} = 3$. Especially the case $N_\mathrm{f} = 3$ is relevant because it describes triplon excitations in dimerized spin systems. We will evaluate sum rules and the thermal occupation in detail to assess the validity of the Br\"uckner approach 
on the level of ladder diagrams at finite temperature.

All results are obtained for the nearest neighbor model in Eq.\ 
\eqref{eq.hamiltonian} without any additional interactions besides the hardcore constraint. We primarily show results for $W/\Delta = 0.5$, 
i.e., for a narrow band. For comparison, we will also show the thermal occupation function in a wide-band case, $W/\Delta = 4$.

\begin{figure}
\centering
\includegraphics[width=1.0\columnwidth]{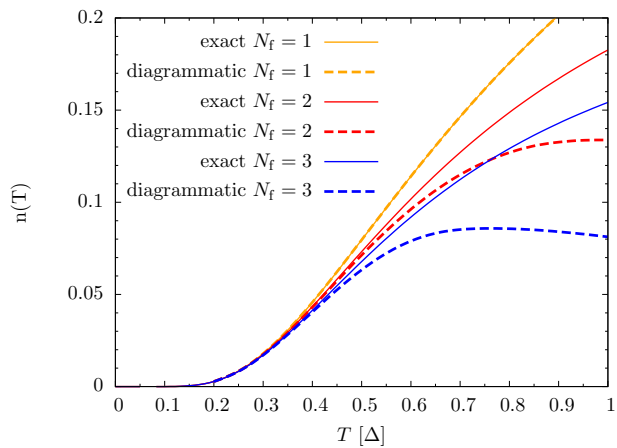}
\caption{Thermal occupation $n(T)$ in the narrow-band 
case $W=0.5\Delta$ for various numbers of flavors $N_\mathrm{f}$ as function of temperature. Comparison between the exact curves (solid lines) and the diagrammatic results (dashed lines).}
\label{fig.thermal_occupation}
\end{figure}

\begin{figure}
\centering
\includegraphics[width=1.0\columnwidth]{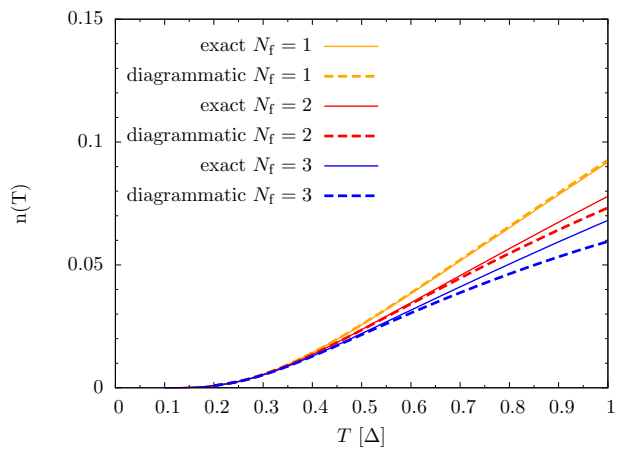}
\caption{Thermal occupation  $n(T)$ in the wide-band case $W=4\Delta$ for various numbers of flavors $N_\mathrm{f}$ as function of temperature. Comparison between the exact curves (solid lines) and the diagrammatic results (dashed lines).}
\label{fig.thermal_occupation_W4}
\end{figure}

First, we study the thermal occupation  which can be computed
from the spectral function by 
\begin{align}
\label{eq.occupation_from_spectral_function}
n(T) = \frac{1}{2\pi} \int\limits_0^{2\pi} \langle b_{k,\alpha}^\dagger 
b_{k,\alpha}^{\phantom\dagger} \rangle \mathrm{d}k =  \int\limits_0^{2\pi} 
\int\limits_{-\infty}^\infty 
\frac{A^\alpha(k,\omega)}{e^{\beta \omega}-1} \mathrm{d} 
\omega \mathrm{d}k.
\end{align}
Note that the thermal occupation is independent of the index 
$\alpha$ because the bosons of different flavor are degenerate.
Figs.\ \ref{fig.thermal_occupation} and \ref{fig.thermal_occupation_W4}
depict the results for different numbers of flavors obtained by the 
Br\"uckner approach and compare them to the exact expression 
\eqref{eq.exact_occupation}.

Since our approach is based on a low-temperature approximation in the expansion parameter $\exp(-\beta\Delta)$ we expect deviations to occur 
if the temperature approaches the value of the energy gap $\Delta$.
Interestingly, the $N_\mathrm{f} = 1$ thermal occupation is
approximated very well even at elevated temperatures \cite{fause14}
and even for vanishing gap \cite{strei15}, only the relative
error diverges for $\Delta\to 0$. This holds for $n(T)$; the deviations 
in the spectral function $A(p,\omega)$ are larger, 
see Section V in Ref.\ \cite{fause14}.

As expected, the diagrammatic and the exact results in Figs.\
\ref{fig.thermal_occupation} and \ref{fig.thermal_occupation_W4}
agree exellently for small temperatures $T \lesssim 0.3 \Delta$. For large temperatures $T\rightarrow \infty$ the exact curves yield the value 
$n(T \rightarrow \infty) = 1/(1+N_\mathrm{f})$ corresponding to the value in
isolated dimers where all states are equally probable.
In contrast to the $N_\mathrm{f} = 1$ case, discernible deviations occur
between the approximate and the exact curves 
for $N_\mathrm{f} > 1$ upon increasing temperature.
The Br\"uckner approach on the level of ladder diagrams 
for $N_\mathrm{f} > 1$ tends to underestimate the thermal occupation at higher temperatures. But we emphasize that for low temperatures the
approach works as expected because it is exact in order $\exp(-\beta\Delta)$, see below.

There is also a significant difference between the performance of the 
ladder approximation for a narrow and for a wide band. At given
gap and temperature the relative deviations are significantly smaller
for wider bands. In the $W=4\Delta$ case, the diagrammatic curves stay close to the exact curves even for $T>0.6\Delta$. In stark contrast, for 
$W=0.5\Delta$, $N_\mathrm{f} = 3$, and $T>0.75\Delta$ the approximate
thermal occupation even shows an unphysical non-monotonic
behavior as function of $T$.

The difference between the two cases can be explained by the different total occupations at the same temperature. For a wide band
the fraction of the Brillouin zone with low-lying excitations in the
range of $T$ is smaller than for a narrow band so that the total
number of thermally excited particles is significantly lower. For this reason, the corrections to the scattering processes considered by the
 ladder diagrams are less important.

This observation implies an important message also for the applicability
of the ladder approximation in higher dimensions. There, the fraction
of the Brillouin zone at low energies for a given gap is smaller 
than in one dimension: Assume the gap to occur at $\vec{k}=0$ then 
\emph{all} components of $\vec{k}$ must be small for
the total energy to be small. For the
same parameters temperature $T$, gap $\Delta$, and band width $W$
one may roughly estimate that the  thermal
occupation in $d$ dimensions $n_d(T)$
scales like $n_1(T)^d$. Thus we expect that for given parameters 
the accuracy of the approximation on the level of
ladder diagrams is significantly higher in higher dimensions.

Next, we will investigate the deviations in one dimension in more detail.
To this end, we study the sum rule of the spectral function 
$A(p,\omega)$. In Eq.\  \eqref{eq.multi_flavor_sum_rule} we saw already  that the integration of $A(p,\omega)$ over frequency $\omega$ is connected to the thermal occupation. Therefore we define the expression
\begin{align}
\label{eq.sum_rule_to_eval}
R[n(T)] := \int\limits_{-\infty}^\infty A^{\alpha}(x) \mathrm{d} x + (1+N_\mathrm{f}) n(T),
\end{align}
which should be equal to unity for the exact spectral function. For brevity, we have introduced  $A(x) = \frac{1}{L} 
\sum\limits_{p} A^{\alpha}(p,x)$, i.e., 
the spectral function averaged over momentum $p$. 

The sum rule \eqref{eq.multi_flavor_sum_rule} states that the integral over the momentum dependent spectral function $A(p,\omega)$ is momentum \emph{independent}. Any approximation to calculate the spectral function is prone to spoil this feature. But we observe in our numerical data
that the Br\"uckner approach completely conserves the exact $p$-independence 
within numerical accuracy for \emph{all numbers} of flavors $N_\mathrm{f}$.
Thus, there is no need to discuss a momentum dependence of the sum
rule. It is sufficient to analyse the quantity $R[n(T)]$.

\begin{figure}
\centering
\includegraphics[width=1.0\columnwidth]{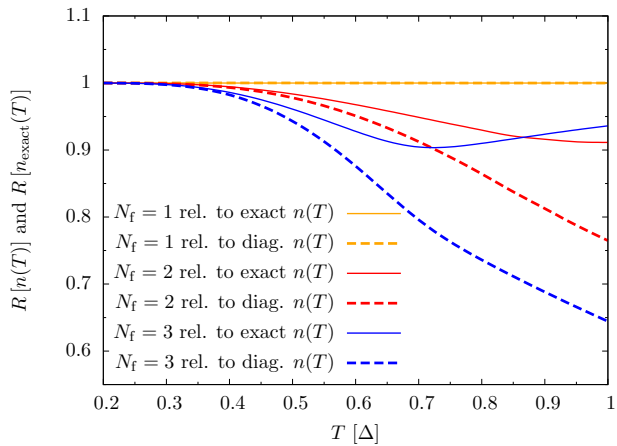}
\caption{The sum rule $R[n(T)]$ from Eq.\ \eqref{eq.sum_rule_to_eval} for various numbers of flavors $N_\mathrm{f}$ as function of temperature. The bandwidth is fixed to $W = 0.5\Delta$.  We compare the quantity $R[n(T)]$ using the diagrammatic $n(T)$ obtained from Eq.\ \eqref{eq.occupation_from_spectral_function} and  
$R[n_\mathrm{exact}(T)]$ using the exact result 
from Eq.\ \eqref{eq.exact_occupation}. }
\label{fig.sum_rule}
\end{figure}

We plot $R[n(T)]$ as function of temperature in Fig.\ \ref{fig.sum_rule}. Since 
for $N_\mathrm{f} > 1$ the thermal occupation $n(T)$ obtained from the spectral function differs from the exact expression in Eq.\ \eqref{eq.exact_occupation}
we depict results for \eqref{eq.sum_rule_to_eval} for both occupation functions, the one determined diagrammatically and the exact one.

For $N_\mathrm{f} = 1$ the sum rule is fulfilled within numerical accuracy
even for  temperatures close to the gap $\Delta$. This changes for higher flavor
numbers $N_\mathrm{f} > 1$. At low temperatures, the sum rule is well fulfilled
because by construction the approach is exact in order $\exp(-\beta\Delta)$.
But deviations become discernible 
at higher temperatures. Some spectral weight seems to be missing
in  the approximate Br\"uckner approach. Note that the sum rule 
turns out to be better fulfilled using the exact $n_\mathrm{exact}(T)$
than using the diagrammatically obtained $n(T)$.
Comparing the case $N_\mathrm{f} = 2$  and $N_\mathrm{f} = 3$, one realizes
that the relative deviation increases upon increasing the number of flavors.

\begin{figure}
\centering
\includegraphics[width=1.0\columnwidth]{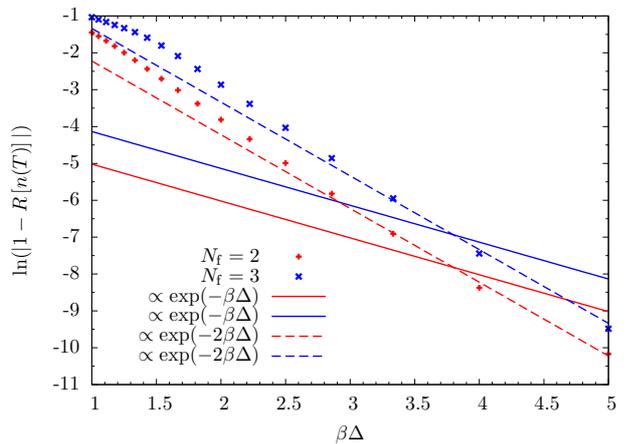}
\caption{Quantitative analysis of the deviation in the calculations of the sum rule for $N_\mathrm{f} > 1$ in Fig.\ \ref{fig.sum_rule} as function of inverse temperature $\beta \Delta$. The bandwidth is fixed to $W = 0.5\Delta$. The $y$-axis shows the deviation of the sum rule from $1$ using the diagrammatic thermal occupation $n(T)$ from Eq.\ \eqref{eq.occupation_from_spectral_function}.  Straight lines indicate exponential dependence of the inverse temperature as expected. The slope indicates the prefactor in the
argument of the exponential.}
\label{fig.sum_rule_error_diagrammatic}
\end{figure}

For a quantitative analysis of the deviations for $N_\mathrm{f}>1$, 
Fig.\  \ref{fig.sum_rule_error_diagrammatic} depicts the logarithmic deviation 
$\ln \left| 1 - R[n(T)] \right| $ of the sum rule in Fig.\ \ref{fig.sum_rule} as function of inverse temperature $\beta \Delta$. We use the diagrammatically obtained occupation function $n(T)$ from 
\eqref{eq.occupation_from_spectral_function}. For low temperatures 
$T \lesssim 0.3\Delta$, i.e., high inverse temperature $\beta \Delta \gtrsim 3.3 $, the deviations smaller than the numerical accuracy of the data

Upon increasing temperature the logarithmic deviation scales linearly with inverse temperature indicating an exponential dependence on inverse temperature. We fitted the functions $A \exp(-\beta \Delta)$ and $B \exp(-2\beta \Delta)$ to the data in the low temperature regime 
$\beta \Delta > 3 $. From these fits it is evident that the deviation indeed scales with $\exp(-2\beta \Delta)$. This in accord with the
construction of the approximation which was designed to be exact in 
linear order in $\exp(-\beta \Delta)$. Thus in the next order, i.e.,
$\propto\exp(-2\beta \Delta)$, deviations cannot be excluded. The analysis
provided underpins the correct reasoning in the choice of diagrams
and in the implementation of the numerical calculations.

\begin{figure}
\centering
\includegraphics[width=1.0\columnwidth]{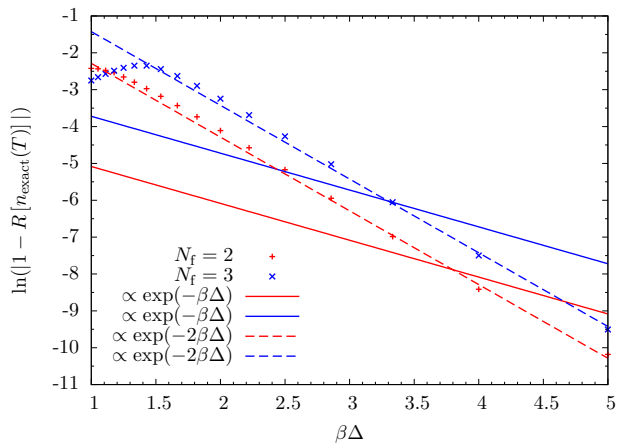}
\caption{Quantitative analysis of the deviation in the calculations of the sum rule for $N_\mathrm{f} > 1$ in Fig.\ \ref{fig.sum_rule} as function of inverse temperature $\beta \Delta$. The bandwidth is fixed to $W = 0.5\Delta$.  The $y$-axis shows the deviation of the sum rule from $1$ using the exact thermal occupation $n_\mathrm{exact}(T)$ from Eq.\ \eqref{eq.exact_occupation}. 
Otherwise as Fig.\ \ref{fig.sum_rule_error_diagrammatic}.}
\label{fig.sum_rule_error_exact}
\end{figure}

In Fig.\ \ref{fig.sum_rule_error_exact},  we additionally show the logarithmic deviation of the sum rule using the exact occupation function $n(T)$ from Eq.\ \eqref{eq.exact_occupation}. Again the deviation scales
as $\exp(-2\beta \Delta)$ indicating that the integral over the spectral function is only correct in first order.

\begin{figure}
\centering
\includegraphics[width=1.0\columnwidth]{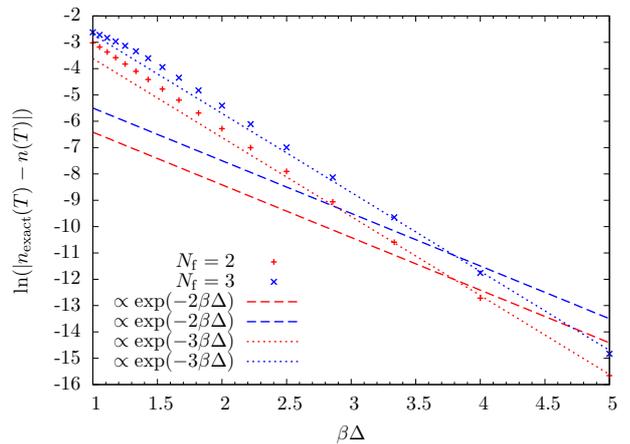}
\caption{Quantitative analysis of the deviations in the calculations of the thermal occupation in Fig.\ \ref{fig.thermal_occupation} as function of inverse temperature $\beta \Delta$ for $N_\mathrm{f} > 1$. The bandwidth is fixed to $W = 0.5\Delta$. The $y$-axis displays the difference between $n_\mathrm{exact}(T)$ from 
Eq.\ \eqref{eq.exact_occupation} and the diagrammatic $n(T)$ from 
Eq.\ \eqref{eq.occupation_from_spectral_function}. Unexpectedly, one finds that the deviations scale proportional to $\exp(-3\beta \Delta)$.}
\label{fig.thermal_occupation_error}
\end{figure}

Next, we compare the thermal occupation  obtained from the 
ladder approximation in Eq.\ \eqref{eq.occupation_from_spectral_function} to the 
exact expression from Eq.\ \eqref{eq.exact_occupation} in Fig.\ 
\ref{fig.thermal_occupation_error}. Again we show the logarithmic difference of the two expressions as function of the inverse temperature. Surprisingly, the difference scales like $\exp(-3\beta \Delta)$.
This indicates a result better than one could expect naively from
the construction of the ladder approximation. 

We attribute  this better-than-expected agreement to the additional averaging of the spectral function over the Bose function in Eq.\ \eqref{eq.occupation_from_spectral_function}. This appears to correct the second order deviation $\exp(-2\beta \Delta)$. We refer the reader to 
the analogy in leading order: The zero-temperature spectral function
corresponds to the zeroth order $\exp(-0\beta \Delta)$ and
leads via the Bose function to the correct result for the thermal occupation in first order $\exp (-\beta \Delta)$. Thus one can
expect that the spectral function in first order implies the correct
thermal occupation in second order.

\section{Analysis at Finite $U$}
\label{sec.finite_U}

In the previous section, we saw that for $N_\mathrm{f} = 1$ the sum rule 
is conserved within numerical accuracy and the thermal occupation 
is approximated  very well by the Br\"uckner approach based on
ladder diagrams. In contrast, the multi-flavor case represents
a less powerful approximation. But we verified that the occurring deviations
are consistent with the systematic construction of the approximation 
as an expansion in powers of $\exp(-\beta\Delta)$. At low temperatures,
the approach works as expected.

Still, these findings are puzzling in view of the concept
of conserving approximations introduced by Baym and Kadanoff
\cite{baym61,baym62} for normal bosons and fermions. 
If the choice of diagrams yields
a conserving approximation for $N_\mathrm{f}=1$ it is
unexpected that this is no longer the case for $N_\mathrm{f}>1$.
We stated, however, already above that the case of hardcore
bosons is more subtle anyway.
To our knowledge, there is no  general argument to prove
that a set of diagrams yielding a conserving approximation for normal bosons
also implies a conserving approximation for hardcore bosons
due to the non-trivial sequence of integration over all frequencies
and the limit $U\to\infty$.

The fundamental question is whether the concept of conserving approximations is generally transferable from normal bosons to hardcore bosons,
i.e., independent of the flavor number $N_\mathrm{f}$.
To understand this issue better and to verify that our
choice of diagrams  provides a conserving approximation
for normal bosons we analyze the approximation of ladder diagrams
for finite repulsion $U$. For finite repulsion, the particles considered are normal strongly interacting bosons. Here the Baym-Kadanoff
concept of conserving approximations is applicable. We study
how the sum rules behave in this case and upon the limit $U\to\infty$.

\begin{figure}
\centering
\includegraphics[width=1.0\columnwidth]{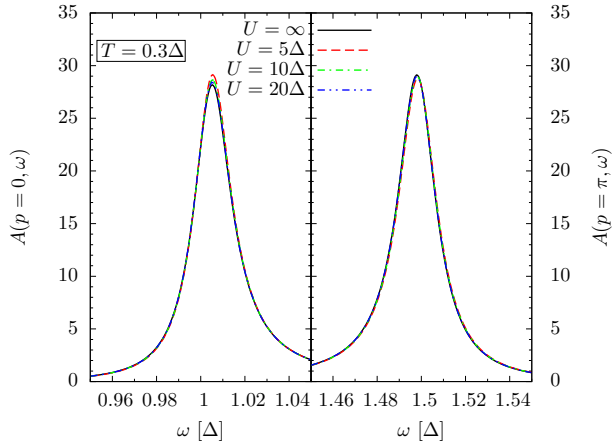}
\caption{Comparison of the spectral lines $A(p, \omega)$ at the gap mode $p=0$ and at the maximum mode $p=\pi$ for various values of the on-site repulsion $U$. The temperature is fixed to $T=0.3\Delta$ and the number of flavors is 
$N_\mathrm{f} = 3$.}
\label{fig.comparison_1}
\end{figure}

\begin{figure}
\centering
\includegraphics[width=1.0\columnwidth]{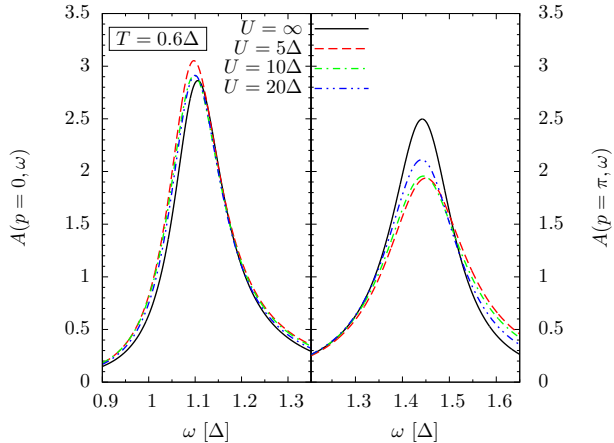}
\caption{Comparison of the spectral lines $A(p, \omega)$ at the gap mode $p=0$ and at the maximum mode $p=\pi$ for various values of the on-site repulsion $U$. The temperature is fixed to $T=0.6\Delta$ and the number of flavors is 
$N_\mathrm{f} = 3$.}
\label{fig.comparison_2}
\end{figure}

In Figs.\ \ref{fig.comparison_1} and \ref{fig.comparison_2},
we compare the spectral line of the scattered boson
for $U=\infty$ with results for 
$U=5\Delta$, $U=10\Delta$ and $U=20\Delta$ . 
For low temperatures, already the case $U=5\Delta$ approximates the $U=\infty$ result quite well. Upon increasing temperature larger deviations
occur. But clearly,  the data for finite values of $U$ 
slowly converge towards to the results at  $U=\infty$. 

For bosons with finite on-site interaction the general sum rule in Eq.\ 
\eqref{eq.sum_rule_general} yields unity for all temperatures
\begin{align}
\int\limits_{-\infty}^\infty A_\mathrm{Bosonic}^{\alpha}(p, \omega) \mathrm{d} \omega = 1.
\end{align}
We carefully checked that the approximation based on ladder diagrams 
indeed fulfills this sum rule independent of the number of flavors 
$N_\mathrm{f}$ and of the temperatures $T$ within numerical accuracy.
Since the spectral lines at finite $U$ converge to the result at 
$U=\infty$ on increasing $U$, there must be additional weight somewhere
in the spectral function to explain the difference to the 
spectral weight in the $U=\infty$ case.

\begin{figure}
\centering
\includegraphics[width=1.0\columnwidth]{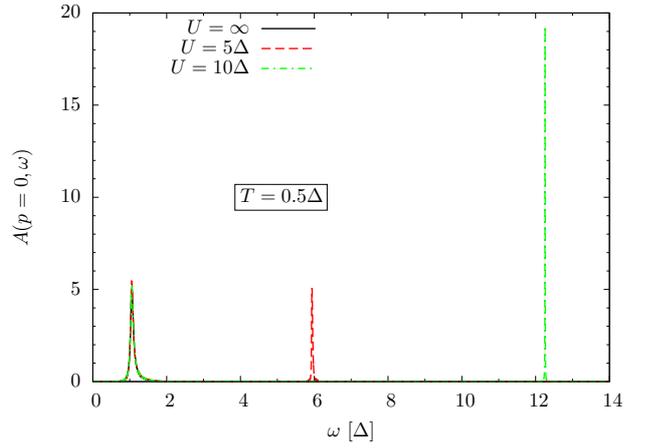}
\caption{Full frequency range comparison of the spectral function $A(p, \omega)$ 
of the gap mode $p=0$ for various values of the on-site repulsion $U$. The temperature is fixed to $T=0.5\Delta$ and the number of flavors is 
$N_\mathrm{f} = 3$. At negative frequencies there is only negligible weight for
 the considered parameters.}
\label{fig.high_frequency1}
\end{figure}

We find the additional weight at energies in the range 
$\omega \approx U$. Fig.\ \ref{fig.high_frequency1} displays the spectral
 function $A(p, \omega)$ over the whole energy range. For finite $U$, the additional weight at high energies can be explained by the formation of an 
anti-bound state of two bosons which repel each other on the same site. 
The signature of this anti-bound state is found in the spectral
function of the scattering amplitude resulting from the Bethe-Salpeter equation 
(not shown). Clearly, the Bethe-Salpeter equation
describes the repeated scattering of two propagating particles.
The energy of the anti-bound state is of the order of the repulsion $U$.

The additional convolution with the third propagator 
only broadens the line by about the band width $W$. 
This is what is seen in Fig.\ \ref{fig.high_frequency1}
at high energies. Note that the position of this
high-energy feature scales with $U$. It is this feature which contains the missing spectral weight for the total sum rule of the single-particle spectral function of normal bosons.

The first key observation for general $N_\mathrm{f}$
 in the above data is the confirmation
that the two limits $U\rightarrow\infty$ of the Br\"uckner approach and $\omega \rightarrow \infty$ in the integration over the frequencies do not commute. The properties of the hardcore bosons are retrieved if
the limit of infinite repulsion is performed first.

For the finite $U$ case, the ladder diagrams represent a conserving approximation independent of $N_\mathrm{f}$. In the infinite $U$ case, the approximation {appears to be a} 
conserving approximation only in the single-flavor 
case $N_\mathrm{f} = 1$, but not for $N_\mathrm{f} > 1$.

We stress that this does not contradict the results by Baym
and Kadanoff, because they formulated conserving approximations for normal
bosonic systems with arbitrary, but finite interactions \tcr{\cite{kadanoffbaym62}. They also discussed that infinite interactions can
be treated by the T-matrix approximation, but did not
consider the subtleties of the sequence of limits, i.e., they did
not elucidate whether this approximation is conserving
for infinite interactions.} 
The single flavor case suggested that the 
Baym-Kadanoff construction also yields conserving approximations 
for hardcore bosons. But our calculations for the $N_\mathrm{f} > 1$ 
falsify this hypothesis. From this observation we deduce our second key conclusion
that  the concept of conserving approximations in the sense of Baym
and Kadanoff does not carry over to hardcore bosons in general.

{The fundamental questions remains why hardcore bosons with multiple flavors behave differently from single-flavor hardcore bosons in one dimension. 
Answering this question is beyond the scope of this article although we
formulate a hypothesis in the Conclusions \ref{sec.conclusion}, 
but further studies are called for.}

\section{Conclusions}
\label{sec.conclusion}

In this paper, we extended the Br\"uckner approach from the 
single-flavor case to the multi-flavor case of hardcore bosons. 
Such multi-flavored excitations are for instance experimentally relevant in coupled 
spin dimers systems where the excitations are threefold degenerate triplons. We
found that the required modifications in the ladder diagrams
for degenerate flavors are minimal. Only the prefactor in the self-energy is modified. 

{The focus of the present work is to gauge the diagrammatic
approach in the multi-flavor case against exact results
and to study conceptual issues concerning conserving approximations
for hardcore bosons.}

{Rigorous results were derived for this purpose for the thermal occupation
of hardcore bosons in the case of a one-dimensional chain with nearest-neighbor hopping. The established rigorous relation is of interest in itself
because it can be used as benchmark for many approximate analytic or
numeric treatments. It could be derived by observing that
the particle degree of freedom and the flavor degree of freedom
represent two independent subsystems for \emph{fixed} number of
bosons. This helps to treat the thermodynamics.
But the dynamic correlations, for instance the hardcore boson propagator,
 cannot be dealt with because they involve changes of the particle number.}

{Our first key results consists in the benchmarks which clearly 
showed that the multi-flavor approach based
on ladder diagrams works very well for low temperatures. This
is expected from the construction of the approach as the leading
linear order of an expansion in powers of $\exp(-\beta\Delta)$.
Thus it can be used up to $T\approx \Delta/2$ for narrow bands and
up to $T\approx \Delta$ for wide bands. For bosons of 
a single flavor it worked even better as established previously
\cite{fause14}. At present, we can only speculate why the
single flavor case is more easily tractable.}

{
The conceptual issue of conserving approximations consists in the
question whether the construction of conserving approximations
by Baym and Kadanoff for normal bosons and fermions \cite{baym61,baym62} carries
over to hardcore bosons. 
In the present article, we confirmed and verified for \emph{finite}
$U$ that our choice of self-consistent ladder diagrams
represents a conserving approximation in the Baym-Kadanoff sense.
This holds true independent of the flavor number $N_\mathrm{f}$.}

{Our previous empirical finding \cite{fause14} in the single-flavor case 
$N_\mathrm{f}=1$, indicated that the hardcore boson sum rules are fulfilled 
\emph{as well} within the numerical accuracy. Thus one could speculate that
the Baym-Kadanoff construction carries over to hardcore bosons.
But the results of the present article for multi-flavor
bosons $N_\mathrm{f}>1$ unequivocally show that this is generally not the case.
This constitutes the second key result of our study.
For several flavors the sum rules are not conserved by the
ladder diagram. We stress, however, that the deviations
only become sizable at higher temperatures.}

We presume that the qualitative difference $N_\mathrm{f}=1$
and $N_\mathrm{f}>1$ is due to
the fact that the case of a single hardcore boson in one dimension
can be exactly mapped to normal fermions for which conserving
approximations exist.  In the case of multiple flavors 
the corresponding Jordan-Wigner mapping to multi-flavor fermions does not
lead to normal fermions because the ensuing fermions are still hardcore
particles repelling each other infinitely strongly if they
are of different flavor. {We can still use this mapping to describe the thermodynamics of multi-flavored hardcore bosons for a fixed number of particles. But dynamic correlations changing the number of particles in the system are not trivially accessible in this way.}

We summarize that the Br\"uckner approach based on self-consistent
ladder diagrams also works for multi-flavor excitations in the interesting low-temperature regime though it is
less {powerful} in comparison to the single-flavor case. 
In particular for wider bands, the approach works very well and we
argued that the approximation should perform even better in higher dimensions,
where the diagrammatic approach can be used as well without
conceptual obstacle.

The presented analysis paves the way for the 
treatment of more realistic models, i.e., models closer
to experimentally relevant systems. Obviously, the dispersion
can be straightforwardly generalized. But also further interactions
beyond the hardcore repulsion need to be included.
The analysis of higher dimensional models is called for {and 
first results in the single-flavor case have become available
very recently \cite{strei15}.}

{Having these extensions in mind, we stress that 
the Br\"uckner approach is a promising tool to describe experimental observations, such as asymmetric thermal line-shape broadening 
as well as thermal narrowing of the single excitation bandwidth.}

Further conceptual extensions concern the addition of the 
considered diagrams to those with \emph{two} backward running
propagators providing an approximation exact up to order 
$\exp(-2\beta\Delta)$.

\begin{acknowledgments}
We thank Piet Brouwer for useful discussions and suggestions. We
acknowledge financial support of the Helmholtz Virtual Institute ``New
states of matter and their excitations''. B.F.\ acknowledges the 
Fakult\"at Physik of the Technische Universit\"at Dortmund for his funding in
the program ``Bestenf\"orderung''. 
\end{acknowledgments}


\end{document}